\documentclass[12pt]{article}
\pdfoutput=1
\usepackage{amsmath}
\usepackage{amsfonts}
\usepackage{setspace}
\usepackage{subcaption}
\usepackage{enumerate}
\usepackage{graphicx}
\usepackage[hidelinks]{hyperref} 
\numberwithin{equation}{section}
\usepackage[utf8]{inputenc}
\newcommand{\gsim}{\lower.7ex\hbox{$\;\stackrel{\textstyle>}{\sim}\;$}}
\newcommand{\lsim}{\lower.7ex\hbox{$\;\stackrel{\textstyle<}{\sim}\;$}}
\def\O{{\mathcal O}}

\newcommand{\be}{\begin{equation}}
\newcommand{\ee}{\end{equation}}
\newcommand{\bea}{\begin{eqnarray}}
\newcommand{\eea}{\end{eqnarray}}

\newcommand{\comment}[1]{}
\newcommand{\expect}[1]{\left\langle #1 \right\rangle}

\newcommand{\bsb}{\boldsymbol}
\newcommand{\Cint}{C\kern-1em\int}

\def\ep{\epsilon}

\def\d{\partial}
\def\vphi{\varphi}

\def\O{\mathcal{O}}

\def\r{{\bsb r}}

\def\tr{{\rm Tr}}

\def\arctanh{{\rm arctanh}}

\def\Im{{\rm Im ~}}

\def\pdot{\dot{\bar\phi}}
\def\P{\mathcal{P}}
\begin{document}
\vspace*{-1. cm}
\begin{center}
{\bf \Large Markovian Dynamics in de Sitter}
\vskip 1cm
{{\bf Mehrdad Mirbabayi} }
\vskip 0.5cm
{\normalsize {\em International Centre for Theoretical Physics, Trieste, Italy}}
\vskip 0.2cm
{\normalsize {\em Stanford Institute for Theoretical Physics, Stanford University,\\ Stanford, CA 94305, USA}}
\end{center}
\vspace{.8cm}
{\noindent \textbf{Abstract:}  
The equilibrium state of fields in the causal wedge of a dS observer is thermal, though realistic observers have only partial access to the state. To them, out-of-equilibrium states of a light scalar field appear to thermalize in a Markovian fashion. We show this by formulating a systematic expansion for tracing out the environment. As an example, we calculate the $O(\lambda)$ correction to the result of Starobinsky and Yokoyama for the relaxation exponents of $\lambda \phi^4$ theory.

\vspace{0.3cm}
\vspace{-1cm}
\vskip 1cm
\section{Introduction}
The goal of this paper is to study the dynamics of a light scalar field from the perspective of a {\em dS observer}. From a different perspective, that of a {\em dS metaobserver}, this problem is over thirty years old, uncovered and largely understood by Starobinsky \cite{Starobinsky}. The metaobserver measures the spatial correlators of the field after the end of inflation. To make predictions for these correlators, interacting fields are usually studied on the expanding background, which after neglecting gravitational dynamics becomes
\be
ds^2 = -dt^2 + e^{2t} dX^2,\qquad \text{(Poincar\'e patch)}.
\ee
But it is also natural to ask what a dS observer sees. This observer lives in de Sitter (almost like ourselves) and describes the spacetime by a static metric
\be\label{static}
ds^2 =- (1-r^2) dt^2 +\frac{dr^2}{1-r^2} + r^2 d\Omega^2,\qquad \text{(static patch)}
\ee
which has a horizon at $r=1$. The question is how the correlation functions decay in time, or equivalently, how an excited state approximately thermalizes due to the migration of perturbations toward the horizon. 

In both pictures, the large excursion of the field plagues a naive perturbative treatment of non-derivative interactions. Even a weak interaction like $\lambda\phi^4$ with $\lambda\ll 1$ has to be treated non-perturbatively for a light enough field. In the extreme case of a massless field, the attractive potential will eventually dominate and stop the random-walk spreading of the free field. 

The canonical way to solve this problem in the Poincar\'e patch is to study the dynamics of the field averaged over a superhorizon region as in figure \ref{fig:slice}-left. Treating this quantity as a stochastic variable, \cite{Starobinsky,SY} derived a Fokker-Planck equation for its probability distribution
\be\label{FP}
\d_t p(t,\vphi) = \frac{1}{8\pi^2}\d_\vphi^2 p(t,\vphi) +\frac{1}{3}\d_\vphi(V'(\vphi) p(t,\vphi)).
\ee
This equation has a ground state (the equilibrium state) and a tower of exponentially decaying eigenmodes that control the approach to equilibrium. In $\lambda \phi^4$ theory, the lowest decay exponent is $\O(\sqrt{\lambda})$, which is indeed a non-perturbative result. Several works, including \cite{Salopek,Tsamis,Finelli,Finelli2,Burgess,Kitamoto,Lopez,Markkanen,Gorbenko,Sundrum,Green,Bounakis,Pinol,Cohen}, have explored various aspects of the problem over the years. Among them \cite{Gorbenko} has been particularly illuminating to the author and answers the Poincar\'e patch version of our main question.
\begin{figure}[t]
\centering
\includegraphics[scale =1.]{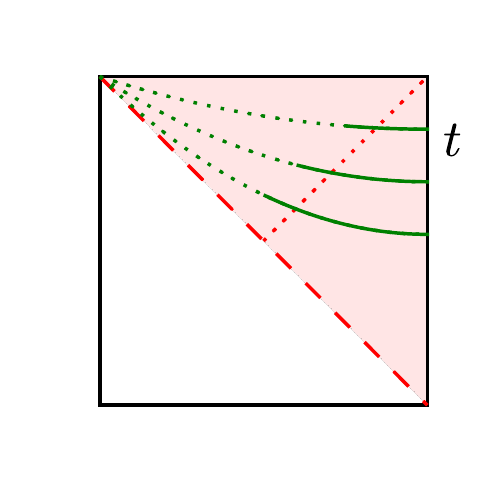} 
~~~~~~~~~~~~~~\includegraphics[scale =1.]{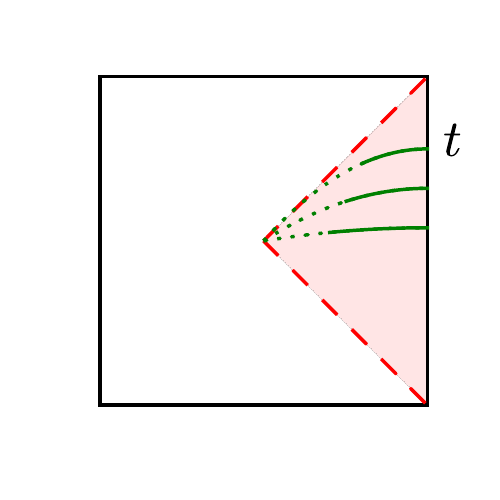} 
\caption{\small{Left: In the stochastic approach of Starobinsky the field is smeared over a superhorizon region of fixed physical size. The Poincar\'e patch is shaded. Right: We study the evolution of $\phi$ smeared over a subhorizon region in the static patch. With an abuse of notation, we have denoted the Poincar\'e patch time and the static patch time both by $t$ even though they are distinct except along the worldline of one observer.}}
\label{fig:slice}
\end{figure}

The analogous observable to study in the static patch is the field smeared over a subhorizon region as in figure \ref{fig:slice}-right. This observable, which we denote by $\bar\phi$, is what a dS observer with access to a finite region would naturally measure. The reduced density matrix of $\bar\phi$ was studied in \cite{infrared}. While the full density matrix $\rho$ evolves unitarily by the static patch Hamiltonian, the evolution of $\rho_{\bar\phi}$ is non-unitary (a natural property of open systems \cite{Agon}). It was shown in \cite{infrared} that the diagonal element of $\rho_{\bar\phi}$ satisfies the same equation \eqref{FP}.

The linearity of equation \eqref{FP} in $V$ is clearly an artifact of it being an approximation; one that we think of as the leading answer in a systematic expansion. A remarkable feature of this leading answer is its Markovianity: the fact that to find $\d_t p(t,\vphi)$ it is enough to know $p(t,\vphi)$ at the same time $t$. It is remarkable because $p(t,\vphi) = \rho_{\bar\phi}(t,\vphi,\vphi)$ is obtained by tracing out a gapless spectrum as reviewed in section \ref{sec:basics}. Therefore, the standard particle physics logic cannot be the justification for the locality of \eqref{FP} in time. More important than explicitly finding the subleading corrections to \eqref{FP} is understanding if they preserve this feature.

Indeed Markovian behavior emerges in dissipative hydrodynamics and Brownian motion in spite of tracing out a gapless spectrum, made up, for instance, of the translational degrees of freedom of the fluid molecules. The hydro modes or floating Brownian particles do interact with this environment, but only through operators whose correlation length is controlled by the mean-free-time and therefore it is much shorter than the time-scales of interest \cite{Nicolis}. We will see in section \ref{sec:marko} that, very much like a Brownian particle, the interactions of $\bar\phi(t)$ with the rest of the system is through operators that, despite including arbitrarily low frequency modes, have an $\O(1)$ correlation time. In fact, we are dealing with a very clean example of the sort: starting from the fundamental theory, one can perturbatively derive the Markovian equation for $p(t,\vphi)$ as an expansion in the ratio of the Markovianity scale to the relaxation time $t_r\gg 1$. 

In section \ref{sec:sub}, we use our perturbative rules to calculate the leading correction to \eqref{FP} and the resulting corrections to the relaxation exponents in $\lambda\phi^4$. We give a general argument (and verify it in the explicit second order calculation) for why these exponents are independent of the precise definition of $\bar\phi$. We conclude in section \ref{sec:con}.

\section{Setup}\label{sec:basics}
This section briefly reviews the essential elements of our setup. More details can be found in \cite{infrared}.
\subsection{Free Spectrum in the Static Patch}
The model consists of a scalar field $\phi$ on the fixed dS geometry
\be
S =\int dt d^3\r\left[-\frac{1}{2} g^{\mu\nu} \d_\mu \phi \d_\nu\phi  + V(\phi)\right]
\ee
where the metric is given in \eqref{static}, and $d^3\r = r^2 dr d^2 \hat r$ with $0\leq r \leq 1$. We diagonalize the kinetic term and treat $V$ as an interaction. Given the spherical symmetry of the background, we expand
\be\label{modes}
\phi(t,\r) = \sum_{l,m} \int_0^\infty \frac{d\omega}{2\pi} \phi_{\omega,l,m}(t)f_{\omega,l}(x)Y_{lm}(\hat r),\qquad x\equiv \arctanh(r).
\ee
As $x\to \infty$ (near-horizon limit), every $l,m$ component of $\phi$ becomes a free $2d$ field. This explains the continuous spectrum of $\omega$. The radial modefunctions $f_{\omega,l}(x)$ are regular at $x=0$, and normalized as\footnote{To make $l=0$ expressions more compact, we normalize $\int d^2\hat r \ Y_{lm}(\hat r) Y_{l'm'}(\hat r) = 4\pi \delta_{ll'}\delta_{mm'}$.} 
\be
\int_0^\infty dx \tanh^2 x \ f_{\omega,l}(x) f_{\omega',l}(x) = \frac{1}{2} \delta(\omega-\omega'),
\ee
so that the free action can be written in the standard form
\be
S_0 = \frac{1}{2}\sum_{l,m}\int_0^\infty \frac{d\omega}{2\pi} \int dt [(\d_t\phi_{\omega,l,m})^2
-\omega^2 (\phi_{\omega,l,m})^2].
\ee
The potential $V(\phi)$ introduces coupling between various $\phi_{\omega,l,m}$. This interaction is localized near the origin $x\sim 1$:
\be\label{sint}
S_{\rm int} = -\int dt dx d^2\hat r \ \frac{\tanh^2x}{\cosh^2 x} \ V(\phi).
\ee
We are interested in the behavior of the correlation functions at large temporal separation, which is controlled by the low frequency modes $\omega\ll 1$. Because of the centrifugal barrier, infrared modes with $l\neq 0$ are suppressed in the interaction region: $f_{\omega,l}(x) \sim \omega x^l$. As a result, at the order we are working, it is enough to know the $l=0$ solution 
\be\label{modefun}
f_{\omega,0}(x) = \sqrt{\frac{1}{\pi(1+\omega^2)}}\left(\cos\omega x +\frac{\omega \sin\omega x}{\tanh x}\right).
\ee
\subsection{A Simple Observable}
We will consider as the observable of interest the smeared field 
\be
\bar\phi(t) = \int \frac{d^3\r}{\sqrt{1-r^2}}  W_\ell(r) \phi(t,\r),
\ee
where $W_\ell$ is a spherically symmetric window function that is centered at the origin and has a characteristic size $r = \tanh \ell$. We can express this just in terms of $\phi_{\omega,0,0}$:
\be
\bar\phi(t) =\int_0^\infty \frac{d\omega}{2\pi} w_\ell(\omega) \phi_{\omega,0,0}(t),\qquad w_\ell(0) = \frac{1}{\sqrt{\pi}},
\ee
where $w_\ell$ is even and analytic near $\omega =0$. The $\omega \sim 0$ behavior of $w_\ell$ follows from the fact that infrared modes are nearly homogeneous in the vicinity of the origin, i.e. $f_{\omega ,0}(x) \approx \frac{1}{\sqrt{\pi}}$ for $\omega \ll 1$ and $x \ll 1/\omega$. 

We expect the relaxation time $t_r$ to be very long, so it is possible (and convenient) to take $1\ll \ell \ll t_r$. Apart from simplifying the computation, this choice suppresses the off-diagonal elements of the near-equilibrium reduced density matrix $\rho_{\bar\phi}(t,\vphi_L,\vphi_R)$ with $|\vphi_L - \vphi_R| \gg 1$. Hence, it allows talking about the classical history of $\bar\phi$ with resolution $\Delta\phi \sim 1$. For our explicit calculations, we take
\be\label{w}
w_\ell(\omega) = \frac{e^{-\omega^2 \ell^2}}{\sqrt{\pi}}\qquad \ell \gg 1,
\ee
and focus on the diagonal element $p(t,\vphi)\equiv \rho_{\bar\phi}(t,\vphi,\vphi)$.
\subsection{In-In Calculus and Its Breakdown}
The equilibrium answer for $\rho_{\bar\phi}$ is the Hartle-Hawking state \cite{HH,GH}, given by the path integral over a cut 4-sphere (i.e. the Euclidean dS$_4$). In particular, saddle-point approximation to this integral gives \cite{infrared}
\be\label{eq}
p_{\rm eq}(\vphi) = a_0 e^{-8\pi^2 V(\vphi)/3}\left(1+\O\left(\frac{V'^2}{V},\frac{V''}{V}\right)\right),
\ee
where $a_0$ is a normalization constant. Our goal is to study the evolution of $p(t,\vphi)$ after perturbing the equilibrium. This would determine the evolution of the equal-time one and multi point correlators of $\bar\phi(t)$ via
\be\label{Odot}
\d_t \expect{\bar\phi^n(t)} = \int d\vphi \vphi^n \d_t p(t,\vphi).
\ee
Alternatively, the knowledge of the left side allows reconstructing $\d_t p(t,\vphi)$. Starting from an initial state $\rho_0$, the in-in perturbation theory is the standard way to calculate these  correlators \cite{Weinberg}
\be\label{commute}
\begin{split}
\expect{O(t)} = \tr(\rho_0 U^\dagger(t) O U(t))=\sum_N i^N\int_0^{t_{N-1}} dt_N\cdots\int_0^{t}dt_1 
\tr(\rho_0 [H_I(t_N),[\cdots,[H_I(t_1),O_I(t)]]\cdots])\end{split}
\ee
where $\expect{.} \equiv \tr(\rho_0 .)$, and the subscript $I$ labels interaction picture quantities. For us $O(t)= \pdot(t) \bar\phi^{n-1}(t) +$ permutations. Of course, the very reason why we focus on the time derivatives, rather than the correlators of just $\bar\phi(t)$, is the posterior knowledge that the evolution is Markovian. 


We take $t\gg 1$ and $\rho_0$ to be a Gaussian state of the interaction picture field, and {\em approximately} thermal for each oscillator $\phi_{\omega,l,m}$ in the expansion of $\phi_I$. If it were the exact thermal state, we would get $\expect{\bar \phi_I^2} = \infty$ since $\phi_I$ is a massless free field with no preferred value. On the other hand, we know that the equilibrium distribution of $\bar\phi$ has a finite variance if $V(\phi)$ is an attractive potential.
So the thermal state of the free theory is infinitely far from that of the interacting theory. One good choice for $\rho_0$ would be to project onto $\bar\phi_I = \vphi_B$ somewhere along the Euclidean time contour that prepares the state, as shown in figure \ref{fig:rho0}. This is an out-of-equilibrium state. More general initial states can be prepared by a weighted sum over $\vphi_B$ (but with a finite range), and by turning on sources. Hence, the interaction Hamiltonian is
\be\label{HI}
H_I(t_i) = \int d^3\r  \ V(\phi_I(t_i,\r)) + s(\bar\phi_I(t_i),t_i),
\ee
where the source $s$ is assumed to turn off well before the time at which we are calculating the correlation functions.
\begin{figure}[t]
\centering
\includegraphics[scale =1.]{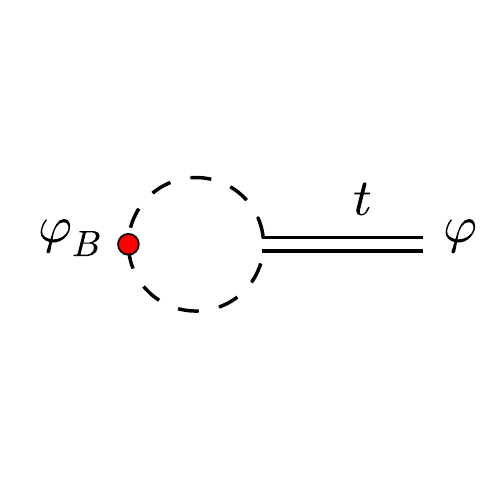} 
\caption{\small{The initial state $\rho_0$ is obtained by the path integral of the free theory over Euclidean dS after projection onto $\bar\phi = \vphi_B$ in the midpoint of the thermal circle. }}
\label{fig:rho0}
\end{figure}

To see why \eqref{commute} is a formal expansion, let us derive an explicit expression for the correlator of $\phi_I$ in $\rho_0$ by analytically continuing the Euclidean result
\be\begin{split}\label{free}
\expect{\phi_I(\tau_1,\r_1)\phi_I(\tau_2,\r_2)} =\frac{1}{Z}&\int D\phi \delta(\bar\phi(\pi)-\vphi_B) e^{-S_0} \phi(\tau_1,\r_1)\phi(\tau_2,\r_2)\\[10pt]
=\frac{1}{Z}&\int\frac{dj}{2\pi} e^{ij\vphi_B}\int D\phi e^{-S_0-ij\bar\phi(\pi)} \phi(\tau_1,\r_1)\phi(\tau_2,\r_2)\\[10pt]
= \expect{\phi_I(\tau_1,\r_1)\phi_I(\tau_2,\r_2)}_\beta& +
\frac{1}{Z}\int\frac{dj}{2\pi} e^{ij\vphi_B}Z (j) \expect{\phi_I(\tau_1,\r_1)}_j\expect{\phi_I(\tau_2,\r_2)}_j.
\end{split}\ee
Here, the path integral is over the thermal circle $0\leq \tau<2\pi$, $Z$ is the partition function (the integral without the $\phi$ insertions), and $Z(j)$ is the partition function of the path integral on the second line. Note that time-translation symmetry is broken by the projection $\bar\phi(\pi) = \vphi_B$.

Expanding $\phi$ as in \eqref{modes}, we have a Gaussian path integral for every $\phi_{\omega,l,m}(\tau)$ with a source term $J_{\omega,l} = \delta_{l,0} j w_\ell(\omega)$ at $\tau = \pi$ and with the Green's function
\be
G_\omega(\tau_1,\tau_2) = \frac{\cosh \omega(\pi+\tau_1-\tau_2)}{2\omega \sinh\omega\pi}.
\ee
The real time correlator is obtained by the analytic continuation $\tau_{1,2} = - it_{1,2}$ with $t_2>t_1$. The thermal contribution, which respects time-translation symmetry, is given by
\be\label{thermal}
\expect{\phi_I(t_1,\r_1)\phi_I(t_2,\r_2)}_\beta =\sum_{l,m} Y_{lm}(\hat r_1)Y_{lm}(\hat r_2)\int_{\tilde\ep} \frac{d\omega}{2\pi} f_{\omega,l}(x_1)f_{\omega,l}(x_2) G_\omega(-it_1,-it_2).
\ee
The lower bound on the $\omega$ integral regulates the IR divergence of the $l=0$ term. This is cancelled by a divergent contribution from the sourced part of \eqref{free}. Up to an additive constant
\be\label{Zj}
\log Z(j) = -\frac{1}{2}j^2 \int_{\tilde\ep} \frac{d\omega}{2\pi} G_\omega(0,0) w_\ell^2(\omega) \equiv - \frac{j^2}{2\ep},
\ee
where $\ep = 4\pi^3 \tilde\ep + \O(\tilde \ep)^2$. The analytically continued 1-point function is
\be\label{phij}
\expect{\phi_I(t,\r)}_j =i j \int_{\tilde \ep}\frac{d\omega}{2\pi} f_{\omega,0}(x) G_\omega(-it,\pi ) w_\ell(\omega) 
= ij \left(\frac{1}{\ep} -c - \frac{t}{8\pi^2}\right)+\O(e^{-t}),
\ee
where the second equality is valid for $x \sim 1$ and $c$ is a constant that depends on $w_\ell$ but its explicit form won't be needed. 
Substituting \eqref{Zj} and \eqref{phij} in \eqref{free}, and neglecting terms that vanish in the limit $\ep \to 0$ or decay exponentially with $t_{1}$ or $t_2$, we obtain the connected correlator
\be\label{free1}
\expect{\phi_I(t_1,\r_1)\phi_I(t_2,\r_2)}_c = \expect{\phi_I(t_1,\r_1)\phi_I(t_2,\r_2)}_\beta -\frac{1}{\ep}+2c+ \frac{t_1+t_2}{8\pi^2}.
\ee
As anticipated, the divergent piece of the thermal correlator \eqref{thermal} is cancelled, but the result is not invariant under time translation. In particular, the equal time correlator grows as $t/4\pi^2$. A closer look (for instance, by integrating in time the equations \eqref{C} and \eqref{A} below) reveals that when $t_2-t_1\gg 1$ the real part of \eqref{free1} is linear in $t_1$ and its imaginary part is $\O(1)$. This leads to the break-down of the in-in expansion \eqref{commute}. Every new term in that expansion comes with one more commutator $i[\phi_I(t_i,\r_i),\phi_I(t_j,\r_j)]\sim 1$ and one time-integral. Therefore, as the total number of $\phi_I$ fields grows with $N$ one faces increasingly faster power-law growth $t^{\frac{1}{2}\#\phi_I}$. For instance, in $\lambda \phi^4$ theory $\#\phi_I\simeq 4 N$ at large $N$. The stochastic method can be thought of as a way to resum this series \cite{Sundrum}.

\section{Markovian Evolution}\label{sec:marko}
In this section we will see that $\bar\phi(t)$ contains the only slow degree of freedom that has to be treated non-perturbatively. We set up a perturbation theory for tracing out the rest.

\subsection{Short-Memory Environment}\label{sec:decay}
Because of the localization of $S_{\rm int}$ near the origin, the free correlators that enter the in-in computation \eqref{commute} can be decomposed into $\bar\phi_I$ correlators and exponentially decaying ones as we will now show.

Let's start by inserting a $\phi_I$ at $\r_1 = {\bsb 0}$ and ask how the correlation with a later insertion depends on position $\r_2$. Only the thermal part of \eqref{free1} depends on the position, and only the $l=0$ in \eqref{thermal} contributes:
\be\label{phiphi}
\expect{\phi_I(t_1,{\bsb 0}) (\phi_I(t_2,\r_2)-\phi_I(t_2,{\bsb 0}))}= 
\int_0^\infty \frac{d\omega}{2\pi} 
\left(\frac{\cos\omega t_{21}}{2\omega\tanh\frac{\beta\omega}{2}}+i\frac{\sin\omega t_{21}}{2\omega}\right)
f_{\omega,0}(0) (f_{\omega,0}(x_2)-f_{\omega,0}(0)),
\ee
where $t_{21} = t_2-t_1$ and $\beta = 2\pi$ in our conventions. A feature of the dS modefunctions $f_{\omega,l}$ is that they are regular functions of $\omega$ at $\omega=0$ with even or odd parity. This is shown in appendix \ref{sec:parity}. Therefore, the integrand of the above integral is even and we can write it as $\int_{-\infty}^\infty d\omega F(\omega) e^{i\omega t_{21}}$, where $F(\omega)$ is analytic near $\omega =0$, and grows as $\omega$ in the limit $\omega\to \infty$. 

The insertion points that are relevant for us are within $x\sim 1$ because the interaction vanishes exponentially for $x\gg 1$. Had we chosen $\r_1 \neq {\bsb 0}$, the UV behavior of the $l=0$ contribution would have been softer, but compensated by the nonzero $l>0$ contribution. The long-time behavior is clearly insensitive to this choice. The $\omega$ integral can be written as
\be
\text{eq. \eqref{phiphi}} = -\d_{t_{21}}^2 \Cint \frac{d\omega}{\omega^2} F(\omega) e^{i\omega t_{21}}
\ee
for a contour that avoids the origin. The double pole at $\omega =0$ results in localized contributions at $t_{21}\sim x_{1},x_2$. Since the singularities of $F(\omega)$ start from $\Im \omega \sim 1$ (after setting $\beta = 2\pi$), when $t_{21}\gg x_1,x_2$ the integral exponentially decays. This would also hold if one of the insertions were smeared to get $\bar\phi_I(t_2)$. We conclude that the correlators of $\phi_I(t,\r) - \bar\phi_I(t)$ are short-range as long as $\arctanh(r)\sim 1$.

The other operator that enters \eqref{commute} is $\pdot(t)$. From \eqref{free1}, we get 
\be\label{beta} 
\expect{\phi_I(t_1,\r)\pdot_I(t_2)}
=\frac{1}{2}\int_0^\infty \frac{d\omega}{2\pi}w_\ell(\omega) f_{\omega,0}(x)
\left(-\frac{\sin\omega t_{21}}{\tanh\frac{\beta\omega}{2}} +i\cos\omega t_{21}\right)
+\frac{1}{4\pi \beta}.
\ee
Indeed, the integral, which comes from the thermal correlator, approaches $-1/4\pi\beta$ exponentially and hence the full correlator decays to $0$. This is guaranteed by the condition $w_\ell(0) = 1/\sqrt{\pi}$ and its analyticity near $0$. For later use, let us evaluate the integral for the explicit choice  of $w_\ell$ in \eqref{w}. The imaginary part, gives
\be\label{C}
[\phi_I(t_1,\r_1),\dot{\bar\phi}_I(t_2)] = \frac{i}{4 \pi^{3/2} \ell}e^{-t_{21}^2/4 \ell^2} + \O(\ell)^{-3}.
\ee
Even though the range of this commutator is $\ell$, as we will see more explicitly, the long time behavior of the system is $\ell$-independent. In particular, if $\ell$ is decreased below $1$, the decay time-scale of \eqref{beta} will be controlled by the singularities of $f_{\omega,0}$ and $\tanh(\pi\omega)$, which are at $\Im \omega = \O(1)$.

The real part of \eqref{beta}, with $w_\ell$ as in \eqref{w}, gives the anti-commutator
\be\label{A}
\expect{\{\phi_I(t_1,\r),\pdot_I(t_2)\}} = -\frac{{\rm Erf}(t_{21}/2\ell)}{2\pi\beta} +\frac{1}{2\pi\beta}+ \O(\ell)^{-2},
\ee
which approaches $0$ as $t_{21}\to \infty$. In fact, the asymptotic behavior of $\expect{\{\bar\phi_I(t_1),\pdot_I(t_2)\}}$ could have been derived just from the knowledge that $\expect{\bar\phi_I^2(t)}\simeq t/2\pi \beta$ at late times.

\subsection{Tracing Out the Environment}
We have identified a hierarchy of time-scales, with $\bar\phi$ being the slow degree of freedom and the one that breaks the standard in-in expansion \eqref{commute}. We can derive a Markovian evolution for $p(t,\vphi)$ by tracing out all other degrees of freedom perturbatively in the inverse powers of the relaxation time 
\be
t_r \sim \expect{\bar\phi^2}_{\rm eq} \gg 1.
\ee
For a given potential, $t_r$ can be related to the coupling constants by noting from \eqref{eq} that at equilibrium $\expect{V} \sim 1$. For instance, $t_r \sim \lambda^{-1/2}$ in $\lambda\phi^4$ theory.

In practice, we first express $\d_t\expect{\bar\phi^n(t)}$ in terms of the equal-time correlators of $\bar\phi(t)$. Explicitly, in the class of perturbed states defined above, and up to corrections that are exponentially small in $t$
\be\label{expand}
\d_t \expect{\bar\phi^n(t)} =\sum_{N=0}^{\infty}\sum_{n_i>0}^{n_0\leq n}
 \frac{n!}{(n-n_0)!}k_{n_0,\cdots,n_N} 
\expect{\bar \phi^{n-n_0}(t) V^{(n_1)}(\bar\phi(t))\cdots V^{(n_N)}(\bar\phi(t))},
\ee
where $n_i = n_0,n_1,\cdots,n_N$, and $k_{\{n_i\}}$ are pure numbers. In our counting, a given term in this expansion is suppressed compared to $\expect{\bar\phi^n(t)}$ by the following power of $1/t_r$:
\be\label{P}
P=\frac{1}{2}\sum_{i=0}^N n_i.
\ee
Therefore, there is a finite number of terms at any order $P$. The existence of the expansion \eqref{expand} follows from a diagrammatic argument. It is based on a separation of variables, similar to how the separation into UV and IR modes in Wilsonian RG makes renormalizablity self-evident \cite{Polchinski}. The key idea is that there are subdiagrams in the in-in perturbation theory in which the fast modes dominate. The various derivatives $V^{(n)}(\bar\phi)$ appear as the effective vertices for these subdiagrams, while every line replaces two $\phi$ fields (an $\O(t)$ quantity in the free theory) with an $\O(1)$ quantity. Hence, the order $P$ of such a subdiagram is simply the number of lines.\footnote{Readers who are impatient to see the results can skip to \eqref{dtp} and proceed to section \ref{sec:sub}.}

\subsubsection*{Standard Diagrammatic Rules}\label{sec:feyn}
We will first briefly review the standard in-in diagrammatics following \cite{Musso}, and then discuss how to modify them to trace out the environment. 
Every in-in diagram consists of trees each representing the time evolution of one of the fields that are correlated. For instance, three tress for $\expect{\bar\phi^2(t)\pdot(t)}$. Each tree combines interaction picture fields (dotted lines) in interaction vertices and evolves them forward with the retarded Green's function (solid lines) toward a future vertex. Hence, there are $n-1$ ingoing lines and one outgoing line at a $\phi^n$ vertex. Abbreviating the coordinates by $a,b$ labels, the retarded function is 
\be\label{GR}
G^{R}_{ab}= i\theta(t_b-t_a) [\phi_{I,a},\phi_{I,b}].
\ee
An example with cubic interaction vertices is shown in figure \ref{fig:diag}-left.
\begin{figure}[t]
\centering
\includegraphics[scale =1]{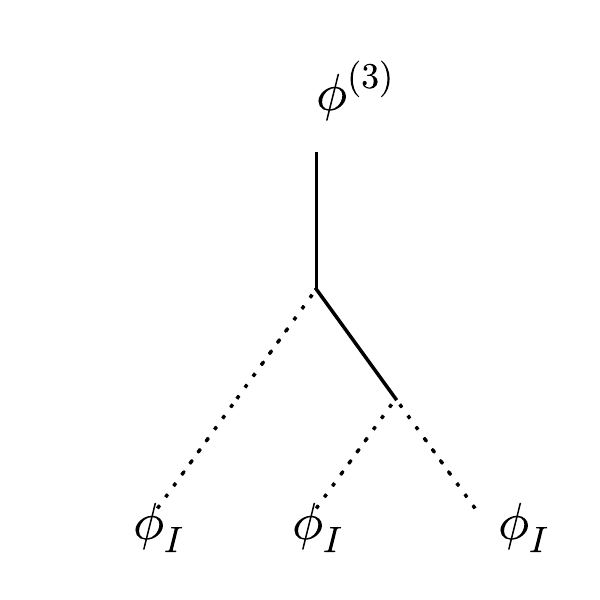} 
~~~~~~~~~\includegraphics[scale =1]{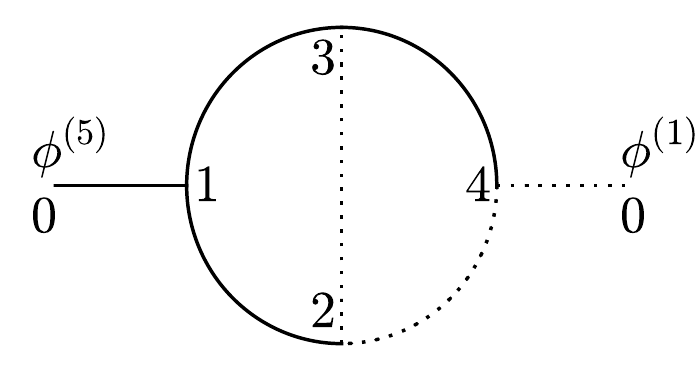} 
\caption{\small{Left: Interaction-picture fields combine in interaction vertices to build a third order field. Right: A contribution to $\expect{\phi^2}$, correlating a fifth order and a first order field. The final insertion can be thought of as the $0^{th}$ vertex. It is the symmetrization vertex of the $0-4$ contraction. The symmetrization vertex for both $2-3$ and $2-4$ lines is vertex $1$.}}
\label{fig:diag}
\end{figure}
In each tree, there is an unambiguous flow of time toward the final field. The trees are glued together by taking the trace in the $\rho_0$ state. If $\rho_0$ is Gaussian, the trace factorizes into free correlators $\expect{\phi_I(t_i,\r_i)\phi_I(t_j,\r_j)}$. After this, we obtain a diagram with external lines ending at the final insertions, and every dotted line representing a free correlator as in figure \ref{fig:diag}-right. The ordering of fields in the free correlators is determined by the set of permutations $\{p_i\}$ of the ingoing lines at every vertex and also all permutations $p_0$ of the indistinguishable final insertions, which can be thought of as the zeroth vertex. Any pair of fields $\phi_{I,a},\phi_{I,b}$ on the two sides of a dotted line has a {\em symmetrization vertex}. This is the vertex at which the time-flows on the two sides of the line first meet (see figure \ref{fig:diag}-right for an example). Suppose this is vertex $i_{ab}$ and the flows meet along the lines $l_a$ and $l_b$ that enter $i_{ab}$. Then we can assign an arbitrary order to the ingoing legs at $i_{ab}$ and for any permutation $p_{i_{ab}}$ of them write
\be\label{Cab}
C_{ab}(p_{i_{ab}})= \theta(p_{i_{ab}}(l_a)-p_{i_{ab}}(l_b))\expect{\phi_{I,a}\phi_{I,b}} +\theta(p_{i_{ab}}(l_b)-p_{i_{ab}}(l_a))\expect{\phi_{I,b}\phi_{I,a}} .
\ee
Note that $\phi_{I,a},\phi_{I,b}$ could also represent the final insertions that are evolved at leading order (e.g. $\phi^{(1)}$ in figure \ref{fig:diag}-right). More generally, the $a,b$ indices include the flavor index if there are multiple fields. After applying this to all dotted lines, we sum over all permutations and multiply by the symmetry factor, i.e one over the number of times these permutations lead to identical assignment of the legs (which are all labeled). Finally, we multiply by the retarded functions and the couplings and integrate every vertex over space and time (starting from the time $\rho_0$ is defined). For a diagram $D$ with $N$ vertices of (possibly the same) order $\{n_1,n_2,\cdots,n_N\}$ and symmetry factor $S(D)$, this procedure gives
\be\label{norm}
S(D)\sum_{\{p_0,p_1,\cdots,p_N\}}\int_{\{t_i,\r_i\}} \prod_{i=1}^N \frac{g_i}{(n_i-1)!}G^R_{a_i a_{i_+}} 
\prod_{\{ab\}} C_{ab}(p_{i_{ab}}),
\ee
where the integral is over the spacetime location of all vertices (with the appropriate measure), an $n^{th}$ order coupling is defined by normalizing the interaction Hamiltonian as $\frac{1}{n!}g\phi^n$, $a_{i_+}$ is the coordinates of the vertex or the final insertion to the immediate future of vertex $i$ (with coordinates $a_i$), and $\prod_{\{ab\}}$ runs over all dotted lines. The combinatorial factors in \eqref{norm} look different from \cite{Musso} because our definition of the symmetry factor is different.
\subsubsection*{Specialized Diagrammatic Rules}
To trace out the short-memory environment, we replace
\be\label{delta}
\phi(t,\r)= \bar\phi(t)+ \delta\phi(t,\r),
\ee
and treat $\bar\phi$ and $\delta\phi$ as two species. The correlation functions of $\delta\phi_I$ with itself and with $\bar\phi_I$ can be calculated explicitly when the smearing function $w_\ell$ is specified. The argument of section \ref{sec:decay} shows that they will all be short-range. In diagrams, we put $\delta$ near the end of a line to indicate $\delta\phi$ on that end. $\bar\phi$ will also be indicated, but no sign means $\bar\phi +\delta\phi$. 

Due to the short range of $\delta\phi_I$ and $\pdot_I$ correlators, there are diagrams with clusters of vertices that are bound together, while different clusters can be separated in time without paying an exponential suppression. When calculating $\d_t \expect{\bar\phi^n(t)}$, there is always one such cluster attached to $\pdot(t)$. Our goal is to treat the dynamics of $\bar\phi$ non-perturbatively. Hence, we can focus on these clusters and instead of long-range correlators of $\bar\phi_I$ that connect them, keep explicit factors of $\bar\phi_I$. We can then ignore all but the cluster that is {\em connected} to $\pdot(t)$ (denoted as the $\pdot$-cluster). The rest are taken into account by promoting every $\bar\phi_I$ into $\bar\phi$ at the end of our calculation.

As such, the vertices in any cluster can have uncontracted $\bar\phi$ fields contained in the time-dependent ``vertex functions'' $V^{(n_i)}(\bar\phi(t_i))$ with $n_i\geq 1$. So the result is not a number, but an unequal-time correlator. 
To keep track of the ordering of the uncontracted fields, we introduce an extra ingoing leg (ending at a cross) at each vertex. Hence in constructing the $\pdot$-clusters, we can have vertices of any order $n$ as long as $V^{(n)} \neq 0$. Since the perturbative expansion is in the number of lines, a non-polynomial $V$ would be as good as a polynomial one. 

The defining property of the cluster is that vertices are bound together. Moreover, dotted lines with $\bar\phi_I$ on both ends are excluded since they are large in perturbation theory. However, because of the causal structure of in-in diagrams solid lines with $\bar\phi$ on both ends are allowed. Figure \ref{fig:bond} shows an example. This is indeed in agreement with our perturbative counting: The retarded function of $\bar\phi$ is $\propto i[\bar\phi_I(t_1),\bar\phi(t_2)]=\O(1)$, so unless it comes with an unrestricted time-integral, it leads to a $1/t_r$ suppression. 
\begin{figure}[t]
\centering
\includegraphics[scale =1]{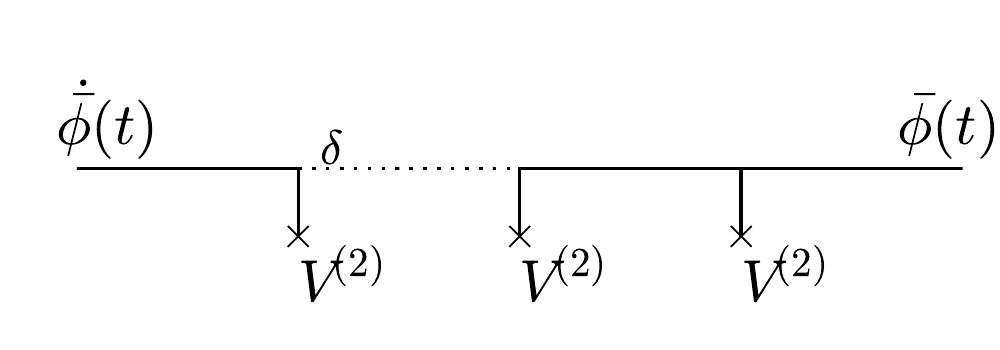} 
\caption{\small{A $4^{th}$ order cluster contributing to $\d_t \expect{\bar\phi^2(t)}$. Label $\delta$ indicates $\delta\phi$ at that end of the line. The two short-range lines on the left and the causal structure of the in-in diagram lead to an exponential suppression if any vertex is taken much earlier than $t$. The crossed legs indicate possible uncontracted $\bar\phi$ fields at the vertices.}}
\label{fig:bond}
\end{figure}

We sum over all $\pdot$-clusters, and within each over all permutations of the ingoing lines (including the crossed legs) into the vertices to obtain
\be\label{time-fold}\begin{split}
\d_t \expect{\bar\phi^n(t)}=\sum_{C}S_n(C)\sum_{\{p_i\}}\int_{\{t_i,\r_i\}}
 \expect{\P\left\{\bar\phi^{n-n_0}(t)
\prod_{i=1}^N \frac{G^R_{a_{i}a_{i_+}}}{n_{\delta\phi,i}! n_{\bar\phi,i}!} V^{(n_i)}(\bar\phi(t_i)) \right\}}
\prod_{\{ab\}} C_{ab}(p_{i_{ab}}).
\end{split}
\ee
Let us compare this with \eqref{norm}. $S_n(C)$ is the symmetry factor associated to $C$, but it depends also on $n$ (more on this below). There is a path ordered expectation value because of the uncontracted fields. The ordering along the Keldysh contour, which is now a time-folded contour like the example in figure \ref{fig:tfold}, is induced by the permutations $p_0,p_1,p_2,\cdots$. The indices $a,b,a_i,a_{i_+}$ include also the flavor information: $(\bar\phi,\delta\phi,\pdot)$, and the factors of $G^R_{ab}$ and $C_{ab}$ exponentially suppress the integrand when any two vertices are separated by $\Delta t\gg 1$. The number of ingoing $\delta\phi$ and $\bar\phi$ legs into the $i^{th}$ vertex are respectively $n_{\delta\phi,i}$ and $n_{\bar\phi,i}$. So we have $n_i = n_{\delta\phi,i} +n_{\bar\phi,i} +1$.

In order to transform \eqref{time-fold} into a sum over equal-time correlators, we Taylor expand
\be\label{phiti}
V^{(n_i)}(\bar\phi(t_i)) = \sum_{m=0}^\infty \frac{1}{m!} V^{(n_i+m)}(\bar\phi(t)) \Big(\int_{t_i}^t dt' \pdot(t')\Big)^m.
\ee
Picking the leading term for all vertices leads to the desired form, with the spacetime-integrals giving a numerical factor. Each higher order term in \eqref{phiti} introduces an extra suppression in $1/t_r$. Hence, at any order $P$ in our perturbative scheme, the expansions can be truncated. The correlation functions involving $\pdot(t')$ can again be written in terms of the time-folded correlation functions of $\bar\phi$ as in \eqref{time-fold}, but now summing over clusters that are connected to $\pdot(t')$. These join the original $\pdot(t)$-cluster to form a bigger cluster of vertices. By repeating the same steps, we arrive at \eqref{expand} after a finite number of iterations for any desired order $P$. 

\begin{figure}[t]
\centering
\includegraphics[scale =1.2]{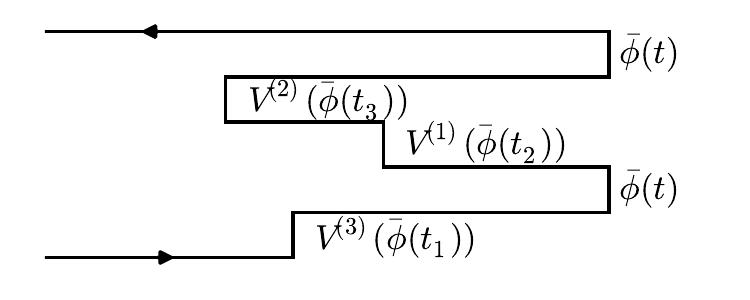} 
\caption{\small{A time-folded correlator $\expect{\bar\phi(t)V^{(2)}(\bar\phi(t_3)) V^{(1)}(\bar\phi(t_2)) \bar\phi(t) V^{(3)}(\bar\phi(t_1))}$ that results from a typical cluster.}}
\label{fig:tfold}
\end{figure}

Lastly, the $n$-dependence of \eqref{expand} follows from the fact that a given cluster $C$ contributes to all $\d_t\expect{\bar\phi^n(t)}$ with $n\geq n_0$, where $n_0$ is the number of final insertions that are connected to the cluster. The symmetry factor would then include a trivial reshuffling of $n-n_0$ {\em idle} insertions, $S_n(C) = S_{n_0}(C)/(n-n_0)!$, and we have
\be
S_n(C) \sum_{p_0} \cdots  = \frac{n!}{(n-n_0)! n_0!}S_{n_0}(C)  \sum_{\hat p_0}\cdots,
\ee
where $\hat p_0$ are the permutations of the $n_0$ fields that are connected to $C$. This produces the correct $n$-dependence that is necessary to translate \eqref{expand} into an equation for $\d_t p(t,\vphi)$ using \eqref{Odot}:
\be\label{dtp}
\d_t p(t,\vphi)=\Gamma_\vphi p(t,\vphi) \equiv \sum_{N=0}^\infty\sum_{n_i>0} k_{n_0,\cdots,n_N} \d_{\vphi}^{n_0}\Big(V^{(n_1)}(\vphi)\cdots V^{(n_N)}(\vphi)\  p(t,\vphi)\Big).
\ee

\section{Thermalization }\label{sec:sub}
Having reduced the problem to a one-dimensional system, we can now solve it non-perturbatively in $\vphi$ to understand its long-time behavior. The discussion of this section parallels that of \cite{SY}, but includes the first correction and comments on the cutoff dependence. 
\subsection{Spectrum of $\Gamma$ and Its Robustness}
The relaxation to the thermal equilibrium as well as the decay of correlators at large separation is governed by the operator $\Gamma$ defined in \eqref{dtp}. To see the latter fact directly, let us consider as an example the correlation function $\expect{\bar\phi_1(t_1)\bar\phi_2(t_2)}$ where the subscripts $1,2$ indicate possibly different smearing lengths $\ell_1,\ell_2$. We can calculate this using a two-point distribution
\be\label{phi21}
\expect{\bar\phi_1(t_1)\bar\phi_2(t_2)}=\int d\vphi_1 d\vphi_2 \ \vphi_1 \vphi_2 \ p(t_1,\vphi_1;t_2,\vphi_2),
\ee
where $p(t_1,\vphi_1;t_2,\vphi_2)$ is obtained by inserting the projector onto $\bar\phi_1=\vphi_1$ and $\bar\phi_2 = \vphi_2$ along the Keldysh contour as in figure \ref{fig:p2}. We can find an equation for $\d_{t_2} p(t_1,\vphi_1;t_2,\vphi_2)$, similarly to $\d_t p(t,\vphi)$ but now by looking at $\d_{t_2}\expect{\bar\phi_1^{n_1}(t_1)\bar\phi_2^{n_2}(t_2)}$. For the same reason that the evolution of $p(t,\vphi)$ is Markovian, the resulting equation reads
\be
\d_{t_2} p(t_1,\vphi_1;t_2,\vphi_2) = \Gamma_{\vphi_2} p(t_1,\vphi_1;t_2,\vphi_2) + \O(e^{-t_{21}}).
\ee
The memory of projection onto $\vphi_1$ gets lost exponentially fast. 
\begin{figure}[t]
\centering
\includegraphics[scale =1.]{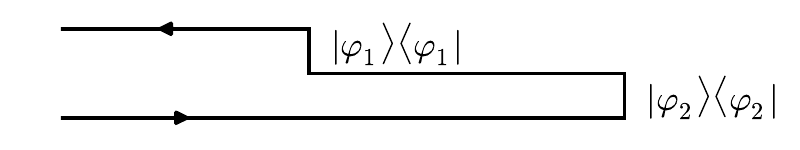} 
\caption{\small{The projection onto $\bar\phi(t_1) =\vphi_1$, $\bar\phi(t_2)=\vphi_2$ to obtain $p(t_1,\vphi_1;t_2,\vphi_2)$.}}
\label{fig:p2}
\end{figure}

At first order in $1/t_r$, $\Gamma$ is given in \eqref{FP}. It is a self-adjoint operator for an appropriate choice of the inner product \cite{SY}. At higher orders, one can perturbatively modify the inner product to obtain a Hermitian operator and find its spectrum $\{\Lambda_n\}$ and eigenstates $\{\Psi_n\}$ using the techniques of time-independent perturbation theory.\footnote{A more detailed analysis at first order can be found in \cite{Markkanen_double}, while the second order case will be carried out below.} Hence, at large $t_{21}$, we can expand
\be\label{eigen}
p(t_1,\vphi_1;t_2,\vphi_2) = \sum_{n} a_n(\vphi_1) \Psi_n(\vphi_2) e^{-\Lambda_n t_{21}}.
\ee
By construction, $\Gamma$ has an overall derivative. This guarantees the conservation of probability and implies $\int d\vphi \Psi_n(\vphi) = 0$ unless $\Lambda_n =0$. For a stable potential the spectrum is non-negative and there is a unique ground state with $\Lambda_0 = 0$. This corresponds to the equilibrium (Hartle-Hawking) state. The nonzero eigenvalues control the long time behavior of the correlator \eqref{phi21} and by a similar argument the relaxation of any simple observable to equilibrium. 

It is perhaps self-evident that the details of how the smeared observables are defined should not affect these exponents. To see it more explicitly, let us take the derivative of \eqref{phi21} with respect to $\ell_2$. We note that the operator
\be
\d_\ell \bar\phi(t) = \int \frac{d\omega}{2\pi} \phi_{\omega,0,0}(t) \d_\ell w_\ell(\omega)
\ee
is similar to $\pdot(t)$ in that its interaction picture correlators are short-range. Hence, following the same steps as in section \ref{sec:feyn}, but replacing $\pdot$-clusters with $\d_{\ell_2} \bar\phi_2$-clusters, we find
\be\begin{split}
\d_{\ell_2}\expect{\bar\phi_1(t_1)\bar\phi_2(t_2)} =\expect{\bar\phi_1(t_1) F(\bar\phi_2(t_2))} 
=\int d\vphi_1 d\vphi_2 \ \vphi_1 F(\vphi_2) \ p(t_1,\vphi_1;t_2,\vphi_2)\end{split}
\ee
for some function $F$ that can be expanded in terms of the sum of products of $V^{(n)}$. The large $t_{21}$ behavior of the right-hand side is controlled by the exponentials in \eqref{eigen}. On the other hand, if the exponents were to depend on $\ell_2$, the left-hand side would have a different time-dependence. To avoid contradiction, we conclude that $\{\Lambda_n\}$ must be independent of the size, and more generally independent of the detailed form of the smearing function as long as it smoothly approaches $\frac{1}{\sqrt{\pi}}$ at $\omega =0$. 

\subsection{Relaxation Exponents at Next-to-Leading Order}
We now focus on the explicit second order equation
\be\label{FPNL}
\d_t p =\Gamma_\vphi p \simeq  \frac{1}{8\pi^2}\d_\vphi^2((1+\alpha_1 V'') p) 
+\frac{1}{3}\d_\vphi((V'+\alpha_2 V'''+\alpha_3 V' V'') p),
\ee
where the coefficients of the subleading terms are
\be\label{alphas}
\alpha_1 = -\frac{16\pi^2}{3} (\alpha_2-c_{\rm UV}) =\frac{2 \ell(2-\sqrt{2})}{3\sqrt{\pi}},\qquad
\alpha_3 = \frac{1}{9}.
\ee
The leading terms correspond to $P=1$ in our counting. As a simple demonstration of our diagrammatic rules, we rederive them. The diffusion term can be obtained from the first order connected contribution to $\d_t\expect{\bar\phi^2(t)}$, shown in figure \ref{fig:first}-left. This is
\be\label{diff}
\d_t\expect{\bar\phi^2(t)}^{(1)}_c= \expect{\{\bar\phi_I(t),\pdot_I(t)\}} = \frac{1}{4\pi^2},
\ee
where we used \eqref{A} for the correlator. 
\begin{figure}[t]
\centering
\includegraphics[scale =1.]{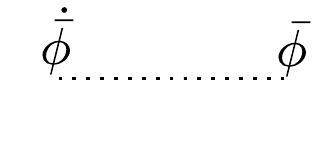} 
~~~~~~~~~~~~\includegraphics[scale =1.]{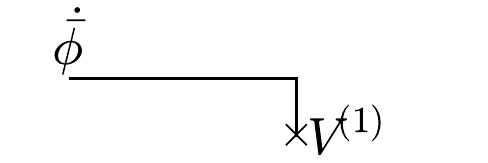} 
\caption{\small{First order diagrams giving the diffusion term (left) and drift term (right).}}
\label{fig:first}
\end{figure}
The leading drift term shows up in the first order contribution to $\d_t\expect{\bar\phi(t)}$, as in figure \ref{fig:first}-right:
\be\label{1pf}\begin{split}
\d_t\expect{\bar\phi(t)}^{(1)}= i \expect{V'(\bar\phi(t))} \int_0^t dt_1\int d^3\r [\phi_I(t_1,\r),\pdot_I(t)] 
= - \frac{1}{3} \expect{V'(\bar\phi(t))},
\end{split}
\ee
where we used \eqref{C} for the commutator, and kept the leading term by substituting $\expect{V'(\bar\phi(t_1))}\to \expect{V'(\bar\phi(t))}$.

The next-to-leading calculation is significantly messier and is deferred to appendix \ref{sec:nlo}. The result \eqref{alphas} is obtained by neglecting $1/\ell^2$ corrections, and it agrees with the finding of \cite{Gorbenko}. There is a UV sensitive contribution $c_{\rm UV}$, which is degenerate with a counter-term. Apart from that, the relation between $\alpha_1$ and $\alpha_2$ has the following implication. If we express \eqref{FPNL} in terms of the ``renormalized'' field
\be\label{phitilde}
\tilde \vphi \equiv \vphi - \frac{1}{2}\alpha_1 V'(\vphi),
\ee
taking into account the transformation of $p$ as a distribution (i.e. $p = \tilde p d \tilde\vphi/d\vphi$), we find $\d_t \tilde p = \tilde \Gamma_{\tilde\vphi} \tilde p$ with an $\ell$-independent $\tilde\Gamma$. Therefore, the relaxation exponents are $\ell$-independent as expected from our general argument.

In fact, after the change of variable the second-order equation takes the same form as the first-order one but with an effective potential
\be
V_{\rm eff} = V +c_{\rm UV} V''+ \frac{1}{18} V'^2.
\ee
Hence, the same steps as in \cite{SY} can be followed to find the eigenvalues and the eigenstates. We can expand the 1-point distribution as
\be\label{Lambdas}
\tilde p(t,\tilde\vphi) = e^{-4\pi V_{\rm eff}(\tilde\vphi)/3}\sum_{n=0}^\infty a_n \Phi_n(\tilde\vphi)e^{-\Lambda_n t},
\ee
where the prefactor defines the appropriate inner product in terms of which $\tilde\Gamma$ is self-adjoint, and
\be\label{schr}
-\Phi_n'' + \left[\left(\frac{4\pi^2 V_{\rm eff}'}{3}\right)^2-\frac{4\pi^2 V_{\rm eff}''}{3}\right]\Phi_n = 8\pi^2 \Lambda_n \Phi_n.
\ee
In particular, $\Lambda_0 = 0$ and 
\be
\Phi_0(\tilde\vphi)= e^{-4\pi^2 V_{\rm eff}(\tilde\vphi)/3},
\ee
giving the equilibrium distribution $\tilde p_{\rm eq.}(\tilde\vphi) = a_0 e^{-8\pi^2 V_{\rm eff}(\tilde\vphi)/3}$. It follows that the equilibrium correlators of $\tilde\phi$ are $\ell$-independent.\footnote{The fact that away from equilibrium they do depend on $\ell$ is shown explicitly in appendix \ref{sec:ell}. Thus, $a_{n}$ with $n\geq 1$ can depend on $\ell$.} Finally, as a concrete example suppose
\be
V(\phi)= \frac{1}{2} m^2\phi^2 + \frac{1}{4}\lambda \phi^4.
\ee
For a light field, it is natural to define the renormalized mass in terms of what appears in the effective potential, $m^2_R = m^2 + 6 \lambda c_{\rm UV}$, since this is the combination that affects well-defined long time observables. For a massless field ($m_R=0$), the first two decay exponents including the first correction to the result of \cite{SY} are obtained by applying time-independent perturbation theory to \eqref{schr}
\be\begin{split}
\Lambda_1 &\approx 1.37\sqrt{\frac{\lambda}{24 \pi^2}} +  1.09  \frac{\lambda}{24 \pi^2},\\[10pt]
\Lambda_2 &\approx 4.45  \sqrt{\frac{\lambda}{24 \pi^2}} + 4.80  \frac{\lambda}{24 \pi^2}.\end{split}
\ee
\section{Discussion}\label{sec:con}
In this paper we studied the thermalization of a light scalar field in de Sitter. We formulated a perturbation theory for deriving the Markovian evolution of $\bar\phi$, the smeared field measured by a dS observer. The key idea was to identify $\bar\phi$ as the only interacting degree of freedom with slow dynamics and to separate it from the environment. This led to a perturbative expansion in powers of $1/t_r$, which is the ratio of two time-scales: the short correlation time of the thermal environment and the long relaxation time of $\bar\phi$. 

In the original approach of Starobinsky, formulated in the Poincar\'e or any expanding patch of dS, Markovianity is expected to arise from the ultralocal evolution of the superhorizon field. This idea, sometimes referred to as the ``separate universe'' picture, has played a ubiquitous role in understanding the evolution of cosmological perturbations (in seminal earlier works such as \cite{Bond,Wands,Creminelli}, as well as more recent applications such as \cite{Fujita,Vennin,Assadullahi,George_pbh,Noorbala,Prokopec,Domcke,George}). The recent work \cite{Gorbenko} shows how this intuition manifests itself in the wavefunction of the universe and ensures Markovianity of the stochastic evolution.

What is the intuition behind Markovianity in the static patch? In a sense, the horizon is the real cause. As we saw, interactions turn off near the horizon, and the system factorizes into a collection of massless free $2d$ fields (one for each spherical harmonic $l,m$). In the tortoise coordinate, in which the horizon is at $x=\infty$, these fields live on the half-line and interact only near the ``boundary'' at $x=0$ (much like the Kondo model \cite{Affleck}). Moreover, the smoothness of the horizon requires a finite temperature $\beta =2\pi$. The thermal correlators of operators made of derivatives of massless free $2d$ fields decay exponentially in time while those of non-derivative composites are long-range. On the other hand, only those operators that are localized near $x=0$ (in a neighborhood with a thickness of the order of the curvature length) are involved in the interactions, excluding all long-range operators except those made of the $l=0$ field at $x=0$. This is the slow degree of freedom captured by $\bar\phi$, and removed from the environment. 

This argument suggests that if we turned on a generic ``bulk'' interaction, one that allowed reflection, we would not obtain a Markovian evolution for $\bar\phi$. It also suggests (supported by the form of correlators in section \ref{sec:decay}) that the scale of Markovianity is determined by the temperature and the curvature length. These two are related in order to have a smooth horizon. However, as a thought experiment, we could keep $\beta$ arbitrary at the expense of a singular horizon and still formulate the question in the static patch (a similar question was raised in \cite{Akhmedov_static}). In particular, if $\beta\gg 1$, we would still find a Markovian evolution but controlled by the larger scale $\beta$. 

It is sometimes argued that the Hartle-Hawking state is unstable due to the particle production or secular effects (see e.g. \cite{Polyakov,Akhmedov,Akhmedov_characters}). Here we do not see any sign of a catastrophe. A large class of perturbed states relax to the thermal equilibrium as far as an observer in dS can see. 

Finally, having understood why the evolution is Markovian, one might wonder how essential it is. For instance, Markovianity ensured that correlation functions can be expanded, as in \eqref{Lambdas}, as a sum of slowly decaying pure exponentials up to corrections that decay at least as fast as $\O(e^{-t})$. This in turn leads to a simple dS-invariant correlator at large separation. But couldn't a similar structure arise from some non-Markovian evolution?

\section*{Acknowledgments}
We thank Luca Delacr\'etaz, Victor Gorbenko, Ignacio Salazar Landea, Alberto Nicolis, Sergey Sibiryakov, and Eva Silverstein for stimulating discussions. This work was partially supported by the Simons Foundation Origins of the Universe program (Modern Inflationary Cosmology collaboration).
\appendix
\section{Parity of the Modefunctions}\label{sec:parity}
For general $l$ the radial modefunctions satisfy
\be
-\frac{1}{\tanh^2 x} (\tanh^2 x f_{\omega,l}'(x))' +\frac{l(l+1)}{\sinh^2 x} f_{\omega,l}(x) = \omega^2 f_{\omega,l}(x).
\ee
They should also be regular at $x=0$. This means that 
\be\label{xto0}
f_{\omega,l}(x\to 0) = A(\omega) x^{l},
\ee
for some function $A(\omega)$ which is fixed by the normalization condition. Let's consider a non-normalized solution $\tilde f_{\omega,l}(x)$ which satisfies \eqref{xto0} but with $A=1$. With this condition $\tilde f_{\omega,l}(x)$ is uniquely fixed. Since the equation depends on $\omega^2$ and the boundary condition is $\omega$ independent, $\tilde f$ is an even function of $\omega$ and regular at $\omega =0$. When $x\gg 1$, it can take one the two asymptotic forms
\be\label{form1}
\frac{A_1(\omega)}{\omega} \sin(\omega(x-x_1(\omega))),
\ee
and
\be\label{form2}
A_2(\omega)\cos(\omega(x-x_2(\omega)))
\ee
where $A_1,A_2,x_1,x_2$ are even functions of $\omega$, and regular at $\omega =0$ (otherwise we could, of course, reduce \eqref{form2} to \eqref{form1}). Any linear combination of these can be rewritten as \eqref{form1}. On the other hand, for the free oscillators to have the standard commutation relation
\be
[a_{l_1m_1}(\omega_1),a^\dagger_{l_2m_2}(\omega_2)]= \delta_{l_1l_2}\delta_{m_1m_2}2\pi\delta(\omega_1-\omega_2),
\ee
the normalized modefunctions $f_{\omega,l}(x)$ have to asymptote to $\frac{1}{\sqrt{\pi}} \sin(\omega x + \delta(\omega))$ for some $\delta(\omega)$. Therefore, 
\be
f_{\omega,l}(x) = \frac{1}{\sqrt{\pi}}\tilde f_{\omega,l}(x)\times\left\{\begin{array}{cc}\frac{\omega}{A_1(\omega)},&\qquad \tilde f \to \text{\eqref{form1}}\\[10pt] \frac{1}{A_2(\omega)},&\qquad \tilde f \to\text{\eqref{form2}}\end{array}\right.,
\ee
which implies the resulting $f_{\omega,l}$ is regular at $\omega =0$ and has a definite parity under $\omega\to -\omega$. In fact, it is easy to see that all but $l=0$ functions are odd. In the $\omega\to 0$ limit, there are two asymptotic forms for $x\gg 1$: constant and linear. The regular solution at $x=0$ goes to a linear combination of the two, which corresponds to the $\omega\to 0$ limit of \eqref{form1}, except for $l=0$, where the regular solution is a constant, which corresponds to \eqref{form2}. This parity cannot change as $\omega$ is continuously increased from $0$.
\section{Next-to-Leading Dynamics}\label{sec:nlo}
In this appendix, we derive the second order equation \eqref{FPNL} for $\d_t p$. The two corrections to the drift term ($\propto  V' V''$ and $ V'''$) can be derived from $\d_t \expect{\bar\phi(t)}$ at second order $1/t_r$. The $V''$ correction to the diffusion term first shows up in the second order connected contribution to $\d_t\expect{\bar\phi^2(t)}$.

\subsection*{(a) Next-to-Leading Drift $\propto V'V''$}
This second order correction, which is shown diagrammatically in figure \ref{fig:a}, is
\be\label{1pf2}\begin{split}
\d_t&\expect{\bar\phi(t)}^{(2a)}=- \expect{V''(\bar\phi(t))V'(\bar\phi(t))} 
\int^t_0 dt_1 \int d^3\r_1 
[\phi_I(t_1,\r_1),\pdot_I(t)]\\[10pt]
\times&\Big[\int_0^{t_1}dt_2 \int d^3\r_2 [\phi_I(t_2,\r_2),\delta\phi_I(t_1,\r_1)]
-\int_{t_1}^t dt' \int_0^{t'} dt_2\int d^3\r_2 [\phi_I(t_2,\r_2),\pdot_I(t')]\Big],\end{split}
\ee
\begin{figure}[t]
\centering
\includegraphics[scale =1.]{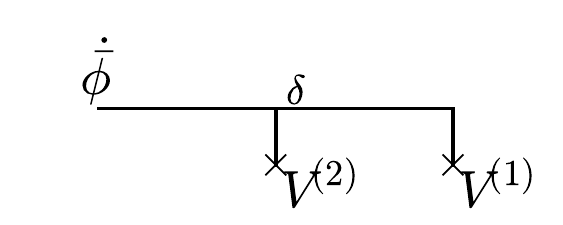} 
\caption{\small{The second order diagram giving the $V'V''$ correction.}}
\label{fig:a}
\end{figure}
where $\delta\phi$ in the first term is defined in \eqref{delta}. The second term arises from $m=1$ in the expansion \eqref{phiti} of the lower order $\pdot$-cluster of figure \ref{fig:first}-right. Strictly speaking, instead of showing both contributions by the diagram of figure \ref{fig:a}, we should have invented another representation for this second term such that the starting point of the second line, which is $\pdot_I(t')$, is somewhere between the two ends of the first line, $t_1<t'<t$. At least at this order, this seems to add more clutter than clarity. 

We can perform the $t'$ integral, and reshuffle terms to get
\be\label{V'1}\begin{split}
\d_t&\expect{\bar\phi(t)}^{(2a)}=- \expect{V''(\bar\phi(t))V'(\bar\phi(t))} 
\int_0^{t_1}dt_2  \int d^3\r_2 \\[10pt]
&\times\Big(
\int^t_0 dt_1 \int d^3\r_1 
[\phi_I(t_1,\r_1),\pdot_I(t)][\phi_I(t_2,\r_2),\phi_I(t_1,\r_1)]-\frac{i}{3} [\phi_I(t_2,\r_2),\bar\phi_I(t)]\Big)\end{split}
\ee
This is equivalent to subtracting from the second order in-in contribution, what is already included in the nonlinear evolution of the first order drift term. We calculate them separately and take the difference. Each calculation is wrong when $t-t_2\gg 1$, but that region cancels in the difference. 

Let's focus first on the first term in \eqref{V'1}. The $\hat r_2$ integral gives
\be\label{r2hat}
\int d\hat r_2 [\phi_I(t_2,\r_2),\phi_I(t_1,\r_1)] = 4\pi i \int \frac{d\omega}{2\pi}\frac{\sin(\omega t_{12})}{\omega} f_{\omega,0}(x_1)f_{\omega,0}(x_2)
\ee
where $t_{12} =t_1 -t_2$, $f_{\omega,0}$ is give in \eqref{modefun} and $r=\tanh x$. Using
\be
\int_0^1 dr_2 r_2^2 f_{\omega,0}(x_2) = \frac{\omega\sqrt{\pi(1+\omega^2)}}{6 \sinh(\pi\omega/2)},
\ee 
we can write
\be\begin{split}\label{r2int}
\int_0^1 d r_2 r_2^2\ \text{\eqref{r2hat}}&=4\pi i(1-\coth(x_1) \d_{x_1})\int \frac{d\omega}{2\pi}\frac{\sin(\omega(t_{12}+x_1))+\sin(\omega(t_{12}-x_1))}{12 \sinh(\pi \omega/2)}\\[10pt]
&= \frac{i}{6} (1-\coth(x_1) \d_{x_1}) (\tanh(t_{12}+x_1)+\tanh(t_{12}-x_1)).\end{split}
\ee
Since $[\phi_I(t_1,\r_1),\pdot_I(t)]$ is independent of $\r_1$ at leading order in $1/\ell$, see \eqref{C}, we take the $\r_1$ integral to get
\be\label{C12}
\int d^3 \r_1 \ \text{\eqref{r2int}}
=\frac{i\pi}{18\sinh^4(t_{12})}(36 t_{12} + 12 t_{12}\cosh(2 t_{12})-26 \sinh(2 t_{12})+\sinh(4 t_{12})).
\ee
This approaches a constant $4\pi i/9$ as $t_{12}\to \infty$ and leads to an unrestricted $t_2$ integral in \eqref{V'1}, but it cancels with a similar contribution from the second term in \eqref{V'1}. In practice, we will perform the full $t_2$ integral and remove $\frac{t}{9} \expect{V''(\bar\phi(t))V'(\bar\phi(t))}$ from both results:
\be
\int_0^{t_1} dt_2 \ \text{\eqref{C12}}
= \frac{4 \pi i}{27}(3 t_1 -1) + \O(e^{-2t_1})
\ee
which after integrating over $t_1$ and using \eqref{C} gives
\be\label{1pf3}
\text{1st term of \eqref{V'1}}_{\rm subtracted}=\left[-\frac{1}{27} -\frac{2 \ell}{9\sqrt{\pi}}\right]
\expect{V'(\bar\phi_I(t)) V''(\bar\phi_I(t))}.
\ee
In the second term in \eqref{V'1} only $l=0$ modes contribute to the commutator:
\be
[\phi_I(t_2,\r_2),\bar\phi_I(t)]=i \int_0^\infty \frac{d\omega}{2\pi} \frac{\sin(\omega (t-t_2))}{\omega}f_{\omega,0}(x_2)w_\ell(\omega) .
\ee
This is independent of $r_2$ at leading order in $1/\ell$:
\be\label{Cphi1}
[\phi_I(t_2,\r_2),\bar\phi_I(t)] = \frac{i {\rm Erf}\left(\frac{t-t_2}{2\ell}\right)}{4\pi} +\O(\ell)^{-2}.
\ee
As anticipated, the $t_2$ kernel approaches a constant $i/4\pi$ when $t-t_2 \gg 1$. Adding what was subtracted in \eqref{1pf3}, we obtain
\be
\text{2nd term of \eqref{V'1}}_{\rm added}
=\frac{2\ell}{9 \sqrt{\pi}}\expect{V'(\bar\phi_I(t))V''(\bar\phi_I(t))}.
\ee
Comparison with \eqref{1pf3} implies the cutoff independent correction $V'V''/27$ to the drift term.

\subsection*{(b) Next-to-leading Drift $\propto V'''$}
The other second order contribution, with the diagrammatic representation in figure \ref{fig:b}, is
\be\label{1pf4}
\d_t\expect{\bar\phi(t)}^{(2b)}=
- \frac{1}{2}\expect{V'''(\bar\phi(t))} \int^t_0 dt_1 \int d\r_1 
\Delta(t,t_1,\r)[\phi_I(t_1,\r),\bar\phi_I(t)],
\ee
\begin{figure}[t]
\centering
\includegraphics[scale =1.]{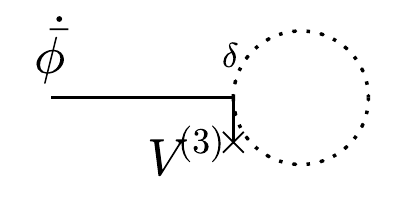} 
~~~~~~~~~~~~~~~~~~\includegraphics[scale =1.]{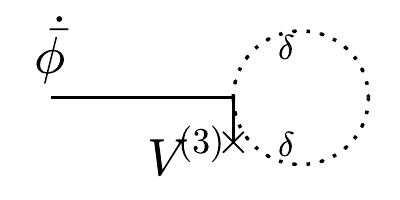} 
\caption{\small{The second order diagrams for the $V'''$ correction. The symmetry factor of the diagram on the right is $1/2$.}}
\label{fig:b}
\end{figure}
where
\be
\Delta(t,t_1,\r) =  \expect{\phi^2_I(t_1,\r)-\bar\phi^2_I(t)}.
\ee
This is obtained by manipulating the four contributions
\be
\expect{\delta\phi^2_I(t_1,\r)}+ \expect{\{\delta\phi_I(t_1,\r),\bar\phi_I(t_1)\}} -
\int_{t_1}^t dt'\expect{\{\bar\phi_I(t),\pdot_I(t')\}} + 
\int_{t_1}^t dt' \int_{t_1}^t dt''\expect{\pdot_I(t')\pdot_I(t'')}
\ee
where the third and forth terms come, respectively, from the $m=1$ and $m=2$ terms in the expansion \eqref{phiti} of the first order $\pdot$-cluster of figure \ref{fig:first}-right.


Using \eqref{free1}, we can write
\be
\Delta (t,t_1,\r) = \frac{t_1-t}{4\pi^2} 
+\int_0^\infty \frac{d\omega}{2\pi} \frac{1}{2\omega \tanh(\pi \omega)}(f_{\omega,0}^2(x)
- w_\ell^2(\omega)) + (l>0).
\ee
The two $l=0$ contributions, $f_{\omega,0}^2$ and $w_\ell^2$, are individually IR divergent. We subtract $1/\pi$ from each to make them IR finite. The integral over $f_{\omega,0}^2 - 1/\pi$ is UV sensitive, but $t$ and $\ell$ independent. Its effect is degenerate with the UV counter-term. The same holds for $l>0$ contribution. So we get
\be
\Delta (t,t_1,\r)= \frac{t_1-t}{4\pi^2} + \frac{\ell}{2\sqrt{2 \pi^5}} + \O(\ell)^{-1}+\text{UV}.
\ee
Substituting this in \eqref{1pf4} gives
\be
\d_t\expect{\bar\phi(t)}^{(2b)}= -\frac{1}{3} \expect{V'''(\bar\phi(t))}
\left(\frac{\ell(\sqrt{2}-2)}{8\pi^{5/2}}+c_{\rm UV}\right) + \O(\ell)^{-1}.
\ee
This adds a correction proportional to $V'''$ to the drift term. 
\subsection*{(c) Next-to-leading Diffusion $\propto V''$}
The second order connected contribution to $\d_t\expect{\bar\phi^2}$ consists of two diagrams, shown in figure \ref{fig:c}, 
\be\label{2pf}\begin{split}
\d_t\expect{\bar\phi^2(t)}^{(2c)}=i \expect{V''(\bar\phi(t))}
\Big[&\int_0^t dt_1 d^3\r \expect{\{\phi_I(t_1,\r),\pdot_I(t)\}} [\phi_I(t_1,\r),\bar\phi_I(t)]
\\[10pt]
+ \int_0^t dt_1 d^3\r & [\phi_I(t_1,\r),\pdot_I(t)]\expect{\{\phi_I(t_1,\r),\bar\phi_I(t)\}
-2 \bar\phi_I^2(t)} \Big]
\end{split}
\ee
where the last two terms arise from 
\begin{figure}[t]
\centering
\includegraphics[scale =1.]{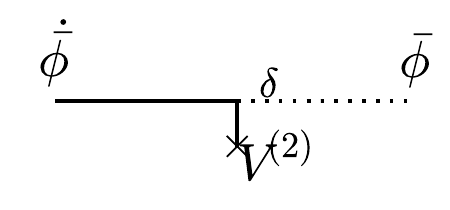} 
~~~~~~~~~~~~~~~~~~~~\includegraphics[scale =1.]{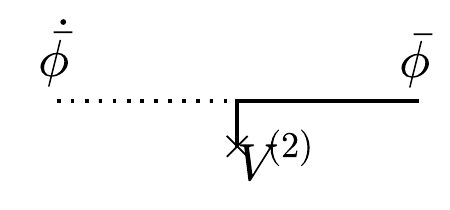} 
\caption{\small{The second order diagrams leading to the $V''$ correction to the diffusion term.}}
\label{fig:c}
\end{figure}
\be
\int_0^t dt_1 d^3\r [\phi_I(t_1,\r),\pdot_I(t)]\Big(
\expect{\{\delta\phi_I(t_1,\r_1),\bar \phi_I(t)\}} -\int_{t_1}^t dt'\expect{\{\bar\phi_I(t),\pdot_I(t')\}}\Big)
\ee
The only new ingredient is 
\be\label{Aphi}\begin{split}
\expect{\{\phi_I(t_1,\r),\bar \phi_I(t)\}}=\expect{\{\phi_I(t_1,\r),\bar \phi_I(t_1)\}}
+\int_{t_1}^{t} dt' \expect{\{\phi_I(t_1,\r),\pdot_I(t')\}}.
\end{split}
\ee
Using the fact that $\d_t \expect{\{\bar\phi_I(t),\bar \phi_I(t)\}} = 1/2\pi^2$, as follows from \eqref{A} and the time-translation symmetry of the thermal state, we can write
\be\begin{split}
\expect{\{\phi_I(t_1,\r),\bar \phi_I(t_1)\}-2\bar\phi_I^2(t)}
&= \frac{t_1-t}{2\pi^2}+
\int_0^\infty \frac{d\omega}{2\pi} \frac{1}{\omega \tanh(\pi \omega)}w_\ell(\omega)(f_{\omega,0}(x)- w_\ell(\omega)) \\[10pt]
&= \frac{t_1-t}{2\pi^2}+\frac{\ell(\sqrt{2}-1)}{2\pi^{5/2}} + \O(\ell)^{-1}.
\end{split}\ee
Using \eqref{A}, we can evaluate the last integral in \eqref{Aphi} to get
\be
\expect{\{\phi_I(t_1,\r),\bar \phi_I(t)\}-2\bar\phi_I^2(t)} 
= \frac{\ell(\sqrt{2}-e^{-(t-t_1)^2/4 \ell^2})}{2 \pi^{5/2}} -(t-t_1) \frac{1+ {\rm Erf}((t-t_1)/2\ell)}{4\pi^2}
+\O(\ell)^{-1}.
\ee
This leads to $\ell \expect{V''}/6 \pi^{5/2}$ from the second line of \eqref{2pf}. In the first line, we use \eqref{A} and \eqref{Cphi1} to get the full result
\be
\d_t\expect{\bar\phi^2(t)}^{(2c)}= \expect{V''(\bar\phi)}
\frac{\ell(2-\sqrt{2})}{6\pi^{5/2}} + \O(\ell)^{-1}.
\ee
This gives a correction $\propto V''$ to the diffusion term. 
\section{Explicit $\ell$-dependence of $\tilde p$}\label{sec:ell}
We have seen that by the field redefinition 
\be\label{phitilde1}
\tilde \vphi \equiv \vphi - \frac{\ell(2-\sqrt{2})}{3\sqrt{\pi}} V'(\vphi),
\ee
the second order evolution equation $\d_t \tilde p =\tilde \Gamma_{\tilde\vphi}\tilde p$ becomes $\ell$-independent. This implies that the eigenvalues and eigenvectors of $\tilde\Gamma$ are $\ell$-independent. However, this does not mean that $\tilde p$ has no explicit dependence on $\ell$. Indeed, the $\ell$ dependence of the original $p(t,\vphi)$ is contained in the field-redefinition \eqref{phitilde1} {\em and} in the coefficients $a_{n>0}$ in the expansion \eqref{Lambdas} of $\tilde p$ in terms these eigenvectors. At the order we are working at, this can be checked for the one and two-point functions. For the explicit choice \eqref{w} of $w_\ell$, the one point function of $\bar\phi$ has the following $\ell$ dependence
\be
\d_\ell \expect{\bar\phi(t)}^{(1)} = i\expect{V'(\bar\phi(t))} \int_0^t dt_1 d^3\r [\phi_I(t_1,\r),\d_\ell\bar\phi(t)]
= \frac{2}{3\sqrt{\pi}}\expect{V'(\bar\phi(t))}.
\ee
Therefore
\be\label{dell1}
\d_\ell \expect{\tilde\phi(t)}^{(1)} = \frac{2}{3\sqrt{2\pi}} \expect{V'}.
\ee
This expectation value is generically nonzero. However, it vanishes at equilibrium $p_{\rm eq.} \propto \exp(-8\pi^2 V/3)$. Note that the keeping next-to-leading corrections to $p_{\rm eq}$ requires working at on higher order in \eqref{dell1}. Furthermore, the connected contribution to $\d_\ell \expect{\bar\phi^2(t)}$ is
\be
\d_\ell\expect{\bar\phi^2(t)}^{(1)}_c = \expect{\{\d_\ell\bar\phi_I(t),\bar\phi_I(t)\}} = -\frac{1}{2\pi^2\sqrt{2 \pi}},
\ee
which combined with \eqref{dell1} gives
\be
\d_\ell\expect{\tilde\phi^2(t)}^{(1)} = -\frac{1}{2\pi^2\sqrt{2 \pi}} +\frac{4}{3\sqrt{2\pi}} \expect{\bar\phi V'}.
\ee
Again, this quantity is generically nonzero but it vanishes at equilibrium.

\bibliography{bibstat}

\end{document}